\documentclass[epj,referee,epsfig]{svjour}
\usepackage{graphicx}
\begin{document}
\title{First and second order ferromagnetic transition at $T=0$ in a 1D 
itinerant system}
\author{S. Daul\cite{address} }

\institute{
Institut de Physique Th\'eorique, Universit\'e de Fribourg,
1700 Fribourg, Switzerland.}
\date{Received: date / Revised version: date}
%
\abstract{
We consider a modified version of the one-dimensional Hubbard
model, the $t_1-t_2$ Hubbard chain, which includes an additional 
next--nearest--neighbor hopping.
It has been shown that at weak coupling this model has a Luttinger liquid 
phase or a spin liquid phase depending upon the ratio of $t_2$ to $t_1$.
Additionally if the on-site interaction $U$ is large enough, the ground state 
is fully polarized. 
Using exact diagonalization and the density-matrix renormalization group,
we show that the transition to the ferromagnetic phase is either of first or 
second order depending on whether the Luttinger liquid or spin liquid is 
being  destabilized.
Since we work at $T=0$, the second order transition is a quantum magnetic 
critical point.
} 
\maketitle
\section{Introduction}
\label{intro}

There has been interest in the theory of zero temperature ferromagnetic transitions
for a few years now. \cite{Kirkpatrick97}
In such transitions, it is the quantum fluctuations, rather than the thermal 
fluctuations, that govern the critical point.
A theory for the onset of ferromagnetism in an unpolarized itinerant system
(Fermi liquid) for dimension $d>1$ was proposed by Hertz \cite{Hertz76}
who showed that the critical behavior should be mean field like.
Precisely for $d=1$, there is no theory for this transition, and
it should be in the same universality class as the onset of ferromagnetism in a 
Luttinger liquid of itinerant electrons. \cite{Sachdev96}
The critical theory of this transition is the main remaining open problem
in the theory of phase transitions in quantum ferromagnetism. \cite{Sachdev96}

In this work, we will present a one-dimensional itinerant model which has 
a ferromagnetic quantum critical point.
We study a modified version of the Hubbard model by including a
next-nearest-neighbor hopping in addition to the nearest one.
The model is no longer integrable, and to investigate it 
we used exact diagonalization and the powerful density
matrix renormalization group (DMRG). \cite{White92}
This model has previously been shown to have a paramagnetic to ferromagnetic 
transition as the on-site interaction $U$ is increased. \cite{DaulNoack98}
Here, we show that the order of the transition is either of
first or second order depending on the parameters of the model.
We will then focus on the second order transition.

\section{Model}

\begin{figure}
\begin{center}
  \includegraphics[width=6cm]{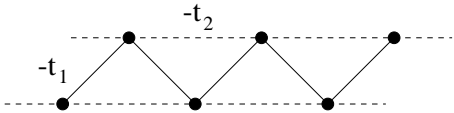}
\end{center}
\caption{The $t_1-t_2$ Hubbard chain.}
\label{figt1t2chain}
\end{figure}

We consider the $t_1-t_2$ Hubbard chain (see Fig.\ref{figt1t2chain})
given by the Hamiltonian 
\begin{eqnarray}
  \lefteqn{ H = -t_1 \sum_{i,\sigma} \left( c^{\dagger}_{i+1\sigma}
                                            c_{i\sigma} + h.c.  
 \right) } \nonumber \\
&&      -t_2\sum_{i,\sigma}\left( c^{\dagger}_{i+2\sigma} c_{i\sigma} 
                                  + h.c. \right)
    + U \sum_i n_{i\uparrow}n_{i\downarrow} . 
\label{eqhamt1t2}
\end{eqnarray}
The summation is over all $L$ sites and spin $\sigma$, and
we will always take $U$ positive.
The sign of $t_1$ is arbitrary since a local gauge transformation,
$c_j \rightarrow e^{i\pi j}c_j$, maps the Hamiltonian with $t_1$ negative
onto the $t_1$ positive Hamiltonian.
Therefore we set $t_1=1$
without loss of generality, and measure all energies in units of $t_1$.
This Hamiltonian conserves the number of particles $N$, the total spin
${\bf S}$ and its projection onto the quantization axis, $S_z$.
If a particle-hole transformation is applied to the system, the transformation
 $t_2 \rightarrow -t_2$ is necessary to recover the original Hamiltonian. 
We restrict ourselves to $0 < N < L$ and $t_2 < 0$, since it is in this
region that a fully polarized ground state (with spin $S=\frac{N}{2}$) 
has been found when $U$ is large enough over a vast region of parameters. 
\cite{DaulNoack98} 
We define $U_c$ as the value of $U$ above which the ground state is
fully polarized.

\begin{figure}
\begin{center}
  \includegraphics[width=5cm]{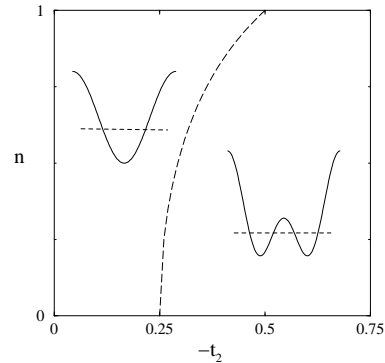}
\end{center}
\caption{The non-interacting phase diagram. The dotted line indicates the boundary
between the systems with two and four Fermi points. }
\label{fig_non_int_phase_diag}
\end{figure}

The weak-coupling phase diagram  has two different regions 
which can be understood by looking at the $U=0$ phase diagram shown in Fig. 
\ref{fig_non_int_phase_diag}. For not too large $|t_2|$, the single-particle
spectrum does not change much from that of the pure Hubbard. 
The Fermi surface has 2 points and we expect the model to be a Luttinger liquid
\cite{Voit} away half-filling.
When $t_2 < t_2^{\mbox{\footnotesize crit}}$, the Fermi surface has 4 points, 
and a weak-coupling treatment predicts that the system should be a spin liquid
(no charge gap but a spin gap). \cite{Fabrizio96}
These two phases have been confirmed with DMRG calculations for 
$U<U_c$. \cite{DaulNoack98}

\section{Order of the transition}

If the system is a Luttinger liquid, it has gapless spin excitations with 
velocity $v_\sigma$.
When it approaches the ferromagnetic transition by increasing $U$, the velocity
$v_\sigma$ might smoothly go to zero  leading to a second order 
transition.
On the other hand, when the system has a spin gap, the transition must be
of first order. To see this, following Ref. \cite{Hertz76}, 
a Hubbard-Stratanovitch transformation is performed on the interacting
term of the Hamiltonian (\ref{eqhamt1t2}). 
The introduced field $m(q,\omega)$\ can be seen as the order parameter, 
and the action of the system is developed in a power series of $m$ as
\begin{equation}
  S[m] = S_0 + \int dq\; d\omega \left[  u_2(q,\omega) m^2(q,\omega)
      + u_4m^4 + \ldots  \right]
\end{equation}
where $S_0$ is the non-interacting action, and $u_n$ is the vertex of order
$n$. 
The quadratic term of the effective action  is
\begin{equation}
  u_2(q,\omega) =  1 - U \chi(q,\omega) 
\end{equation}
with $\chi(q,\omega)$  the spin susceptibility. 
Since
\begin{equation}
      \lim_{q\rightarrow 0} \lim_{\omega\rightarrow 0} \chi(q,\omega) = 0
\end{equation}
for a spin-gapped system then
\begin{equation}
  S[m] = S_0 + \int dq\; d\omega\; m^2(q,\omega) \left[ 1 + 
                      {\cal O} (q^2,\omega^2)  \right] .
\end{equation}
Since the coefficient of $m^2$ does not depend on $U$, the transition must
be of first order.

Numerically we can study the order of the phase transition by calculating
the ground-state energy $E_0(U)$ around $U_c$ with very high precision. 
Since there are many states with energy very close to $E_0$, a
large number of iterations are needed in the Davidson procedure used in exact
diagonalization in order to obtain convergence 
(more than 1000 $H | \psi \rangle$ multiplications).
If the transition is first order, the ground state will jump from $S=0$
to $S=S_{\mbox{\footnotesize{max}}}$, and $E_0(U)$ will have a kink at $U_c$,
since the fully polarized state has no $U$ dependence.
On the other hand, if the transition is second order, 
the energy and spin will smoothly take on all values
from 0 to $S_{\mbox{\footnotesize{max}}}$ as a function of $U$.
In the thermodynamic limit, a second order transition requires that
\begin{equation}
    \lim_{U\rightarrow U_c^-} \frac{\partial E_0}{\partial U} = 
   \lim_{U\rightarrow U_c^+} \frac{\partial E_0}{\partial U},
   \label{eqcontenergy}
\end{equation}
i.e., the derivative of the ground-state energy is continuous
through the transition, while it is discontinuous for a first order
transition.
In order to clarify this issue, we  follow the lowest energy
state with a particular spin $S$. 
However utilizing the ${\bf S}^2$ quantum number in exact
diagonalization is technically difficult, 
and so we follow a state of a particular $S$ by diagonalizing the
augmented Hamiltonian 
\begin{equation}
   H' = H + \lambda {\bf S}^2
\end{equation} 
in different $S_z$-subspaces with $\lambda > 0$. 
For large enough $\lambda$, the lowest energy state within a given
$S_z$ sector will have the minimum $S$ value. \cite{DaulNoack98}

\begin{figure}
  \includegraphics[width=7cm]{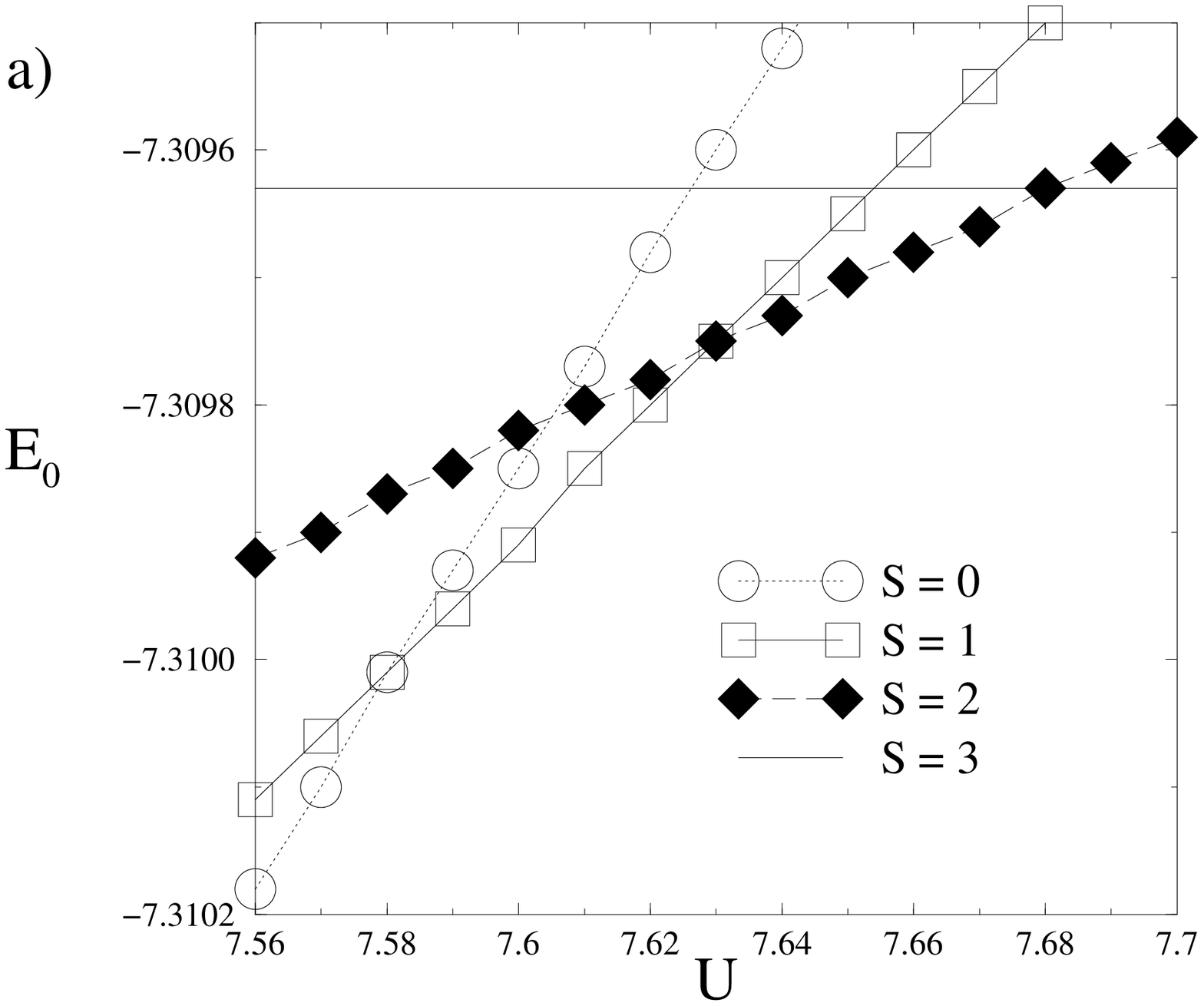}
  \includegraphics[width=7cm]{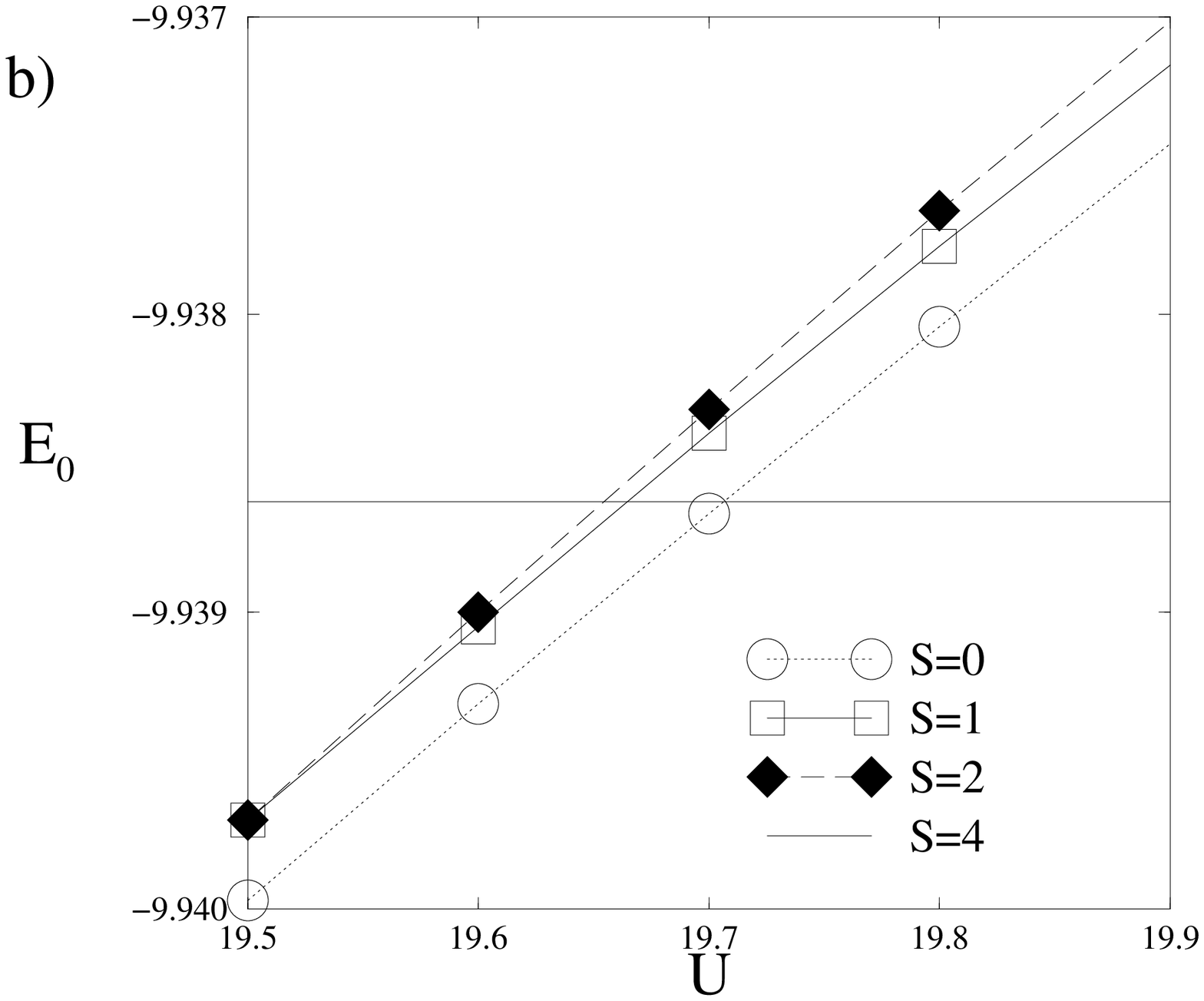}
\caption{Ground-state energy as a function of $U$ for different 
     $S$-subspaces. a) The system has $L=12$, $N=6$ and $t_2=-0.2$.
	b) The system has $L=14$, $N=8$ and $t_2=-0.8$.
    Note that the solid horizontal line is the ground-state energy of the 
     fully polarized state and does not depend on $U$. The state $S=3$ is
    not shown in b) because it is higher in energy.
}
\label{figorder}
\end{figure}

Results obtained with the Davidson algorithm \cite{Davidson} 
for two different cases are shown in Fig. \ref{figorder}. 
In Fig. \ref{figorder} a), for $t_2=-0.2$, $L=12$ and $N=6$, 
when the non-interacting Fermi surface
has two points, we clearly see that the spin $S$ of
the ground state takes on all intermediate values as $U$ is increased.
This is an indication that Eq.\ (\ref{eqcontenergy}) will be satisfied
in the thermodynamic limit and that the transition is continuous.
On the other hand, in Fig. \ref{figorder} b) we see that the transition for a 
system with $t_2=-0.8$, $L=16$ and $N=8$, which is in the spin liquid phase at
weak $U$, is from the $S=0$ state to the fully polarized one with $S=4$,
indicating a first order transition.

\begin{figure}
  \includegraphics[width=7cm]{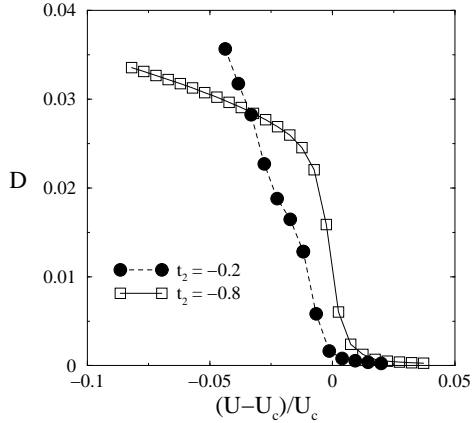}
\caption{Double occupancy $D$ as a function of renormalized $U$ for two 
systems with $L=40$ and $N=20$. The value of $U_c$ is determined as in Ref. 
[5].}
\label{figdocc}
\end{figure}

Another way of determining the order of the transition is to study the first 
derivative of the ground-state energy with respect to $U$ directly. 
For the $t_1-t_2$ Hubbard chain, this is simply the double occupancy
\begin{equation}
  \frac{\partial E_0}{\partial U} = \sum_i \langle n_{i\uparrow}
               n_{i\uparrow} \rangle = D.
\end{equation}
Here we perform DMRG calculations keeping up to 800 states on lattices of up to 80
sites so that the maximum weight of the discarded density matrix
eigenvalues is $10^{-6}$. 
Figure \ref{figdocc} shows the results as a function of renormalized $U$. 
We see that for $t_2 = -0.2$ (full circles) the transition is continuous 
whereas when $t_2=-0.8$ (open squares), it seems discontinuous. 
It is certainly the mixing of energetically close states near the 
transition which makes it seem smooth.

\section{Critical Exponents}

We will now focus on the second order transition. 
Here the relevant scaling field is not the temperature but the interaction 
\begin{equation}
      g = |U-U_c| . 
\end{equation}
The order parameter $m = \langle S_z \rangle $ is coupled to 
the external field $h$.
In addition to classical exponents one has to introduce a dynamical exponent 
$z$. 
The homogeneity hypothesis for the energy and the correlation function are
\cite{homogeneity}
\begin{equation}
      E(bg,b^\beta m) = b^{2-\alpha} E(g,m)
\end{equation}
and
\begin{equation}
      \Gamma( b^{-1/\nu}g, b^{-y}h, br, b^{z}\tau  ) =
         b^{2-(d+z+\eta)}  \Gamma(g,h,r,\tau)
\end{equation}
with $d$ the dimension, here $d=1$.
This leads to the well-known definition of all critical exponents and to
the following identities 
\begin{eqnarray}
 \alpha + 2\beta + \gamma = 2 \\
 \gamma = \nu(2-\eta) \\
 \alpha + \beta(\delta+1) = 2\\
 (d+z) \nu = 2 - \alpha.
\end{eqnarray}
with $y$ related to $\delta$ by $y = (d+z)\delta/(1+\delta)$.

In addition, there is an identity derived by Sachdev. \cite{Sachdev94}
Since the correlation function
\begin{equation}
G(r,\tau) = \langle S_z(r,\tau) S_z(0,0) \rangle
\end{equation}
scales like
\begin{equation}
  G(br,b^z\tau) = b ^{2-(d+z+\eta)} G(r,\tau),
\end{equation}
the scaling dimension $\mu$ of $S_z$ is 
\begin{equation}
 \mu = \frac{d-2+z+\eta}{2}.
\end{equation}
However since $S_z$ is a conserved charge density (it commutes with $H$),
below the upper critical dimension its scaling dimension must be precisely $d$.
Therefore, $\mu = d$ which leads to the identity
\begin{equation}
   z = d + 2 -\eta.
\end{equation}
Thus for $d=1$, one finds that $\nu=\beta$ and $z=1 + \frac{\gamma}{\beta}$.

\begin{figure}
  \includegraphics[width=7cm]{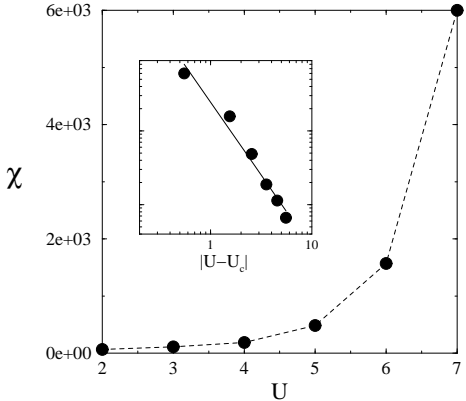}
\caption{The spin susceptibility for $L=60$, $t_2=-0.2$ and $n=0.5$ as a 
function of $U$. The inset shows the points on a log-log scale fitted by
the form Eq. (\protect\ref{gamma_def}) with $\gamma = 1.9$.}
\label{fig_chi}
\end{figure}

The critical exponent $\gamma$ defined by
\begin{equation}
     \chi \sim g^{-\gamma}
\label{gamma_def}
\end{equation}
can be obtained in two different ways.
The first is by taking advantage of the fact that the 
system is a Luttinger liquid, for which the spin susceptibility is inversely 
proportional to the spin velocity. 
This leads for $t_2=-0.2$ and $n=0.5$ to a critical exponent of 
$\gamma = 2.0 \pm 0.1$. \cite{DaulNoack98} 
Another way is by adding to the Hamiltonian an external field $h$ coupled 
to $S_z$ 
\begin{equation}
  H' = H + hS_z,
\end{equation}
where the susceptibility is then given by
\begin{equation}
  \chi = \lim_{h\rightarrow 0} \frac{\partial \langle S_z \rangle}{\partial h}.
\end{equation}
The results for the same parameters as mentioned before are shown in 
Fig. \ref{fig_chi}. 
The critical exponent obtained by a least-square fit is 
$\gamma = 1.9 \pm 0.1$ where the error comes from the fit. 

\begin{figure}
  \includegraphics[width=6cm]{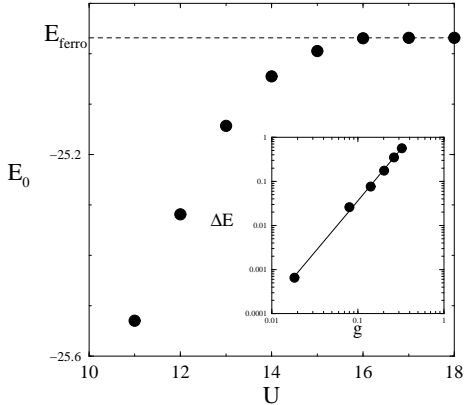}
\caption{The ground-state energy as a function of $U$ for $L=40$, $t_2=-0.15$
and $n=0.6$. The dashed line is the energy of the fully polarized state.
The inset shows the points on a log-log scale fitted by
the form in Eq. (\protect\ref{alpha_prime_def}) with $\alpha' = 2.33$.}
\label{fig_deltaE}
\end{figure}

Another critical exponent can be obtained, namely $\alpha$. Normally this 
exponent is defined by the divergence of the heat capacity. 
Since we deal with a quantum critical point this critical exponent is defined 
by 
\begin{equation}
  \Delta E (U) = E_{\mbox{\footnotesize ferro}} - E_0(U) \sim g^{\alpha'}
\label{alpha_prime_def}
\end{equation}
with $\alpha' = 2 - \alpha$.
Figure \ref{fig_deltaE} shows the results for  $L=40$, $t_2=-0.15$ and $n=0.6$.
A mean square fit yields $\alpha' = 2.33 \pm 0.05$, where the error comes from
the fit. A calculation for $L=80$ yields the same exponent.

Since these two exponents ($\alpha'$ and $\gamma$) only involve the 
evaluation  of ground-state energies, they can be obtained with sufficient 
precision. 
From the identity between exponents, we get a small $\nu$  which is 
consistent with previous calculation of correlation functions. 
\cite{DaulNoack98}
Table I shows the critical exponents obtained for three different
sets of parameters. 
We clearly see that they are not universal, and that $\beta$ is small while 
$z$ is large.
This certainly is the sign of a crossover with the nearby first order transition
(for which $\beta = 0$), due to finite size effects.
The fact that the first set of parameters, $t_2=0.1$ and $n=0.4$, 
which are the farthest from the first order transition, 
gives the largest $\beta$ goes also in the direction of a crossover.

\begin{table}
\caption{Critical exponents for three different set of parameters. 
The errors are from the least-square fit. The critical exponents $\beta$ and
$z$ are calculated using the identities between exponents.}
\[
\begin{array}{|c c|c|c|c|c|c|} \hline
  t_2  &     n        & U_c & \alpha' & \gamma    & \beta    & z  \\ \hline
-0.1 & 0.4  & 15.1 & 1.34 \pm 0.23 & 1.15 \pm 0.16 & 0.1 & 12.5\\
-0.2 & 0.5  & 7.54 & 1.87 \pm 0.03 & 1.83 \pm 0.2  & 0.08 & 24\\
-0.15 & 0.6 & 16.3 & 2.33 \pm 0.05 & 2.17 \pm 0.06 & 0.02 & 100\\ \hline
\end{array} 
\]
\end{table}

\section{Conclusion}

In conclusion, the $t_1-t_2$ Hubbard chain, with negative $t_2$ and filling
less than one half, shows a transition from a spin liquid  or a
Luttinger liquid to a ferromagnet, depending on the ratio $t_2/t_1$.
This transition is of first order when we increase $U$ from the spin
liquid regime and second order when we increase $U$ from the Luttinger liquid.
This second order transition is characterized by a quantum critical point 
for which we can extract exponents.
However the systems considered are relatively small and we see a crossover
with the first order transition due to finite size effects.
More work has to be done by looking at even larger systems and also by 
trying to extract the dynamical exponent $z$ directly.

\section{Acknowledgments}

This work was supported by the Swiss National Foundation.
We would like to thank D.\ Baeriswyl, D.\ Duffy and T.\ Senthil 
for helpful discussions and also R. M. Noack for important help
with the DMRG program.

\end{document}